\documentclass{achemso}
\usepackage[version=3]{mhchem}

\usepackage{graphicx}
\usepackage{dcolumn}
\usepackage{bm, amssymb}
\usepackage[colorlinks=true, allcolors=blue]{hyperref}

\usepackage{tikz,xcolor} 

\definecolor{lime}{HTML}{A6CE39}
\newcommand{\orcidicon}{%
	\begin{tikzpicture}
	\draw[lime, fill=lime] (0,0) 
	circle [radius=0.16] 
	node[white] {{\fontfamily{qag}\selectfont \tiny ID}};
	\draw[white, fill=white] (-0.0625,0.095) 
	circle [radius=0.007];
	\end{tikzpicture}
	\hspace{-2mm}
}

\newcommand{\orcid}[1]{\href{https://orcid.org/#1}{\orcidicon}}

\global\long\def\Re{\operatorname{Re}}
\global\long\def\Im{\operatorname{Im}}

\title{Electric field control of superconducting fluctuations and quasiparticle interference at oxide interfaces}

    \author{Graham Kimbell\orcid{0000-0001-9610-3589}}
    \email{Graham.Kimbell@unige.ch}
    \affiliation[University of Geneva]{Department of Quantum Matter Physics, University of Geneva, Geneva, Switzerland}
    \altaffiliation{These authors contributed equally to this work}
    
    \author{Ulderico Filippozzi\orcid{0000-0001-8146-0006}}
    \affiliation{Kavli Institute of Nanoscience, Delft University of Technology, Delft, Netherlands}
    \altaffiliation{These authors contributed equally to this work}
    
	\author{Stefano Gariglio\orcid{0000-0001-8263-4506}}
	\affiliation[University of Geneva]{Department of Quantum Matter Physics, University of Geneva, Geneva, Switzerland}

    \author{Marc Gabay}
    \affiliation{Laboratoire de Physique des Solides, Universite Paris Saclay, CNRS UMR 8502, Orsay Cedex, France}
	
	\author{Andreas Glatz \orcid{0000-0002-2007-3851}}
    \affiliation{Materials Science Division, Argonne National Laboratory, Argonne, Illinois, USA} 
    \alsoaffiliation{Department of Physics, Northern Illinois University, DeKalb, Illinois, USA}
	
    \author{Andrey Varlamov\orcid{0000-0002-5522-9182}}
	\affiliation{Institute of Superconductivity and Innovative Materials (CNR-SPIN) Rome, Italy}
	\alsoaffiliation{Lombard Institute ``Academy of Sciences and Letters'', via Borgo Nuovo 26, Milan, 20121, Italy}

	\author{Andrea Caviglia\orcid{0000-0001-9650-3371}}
	\email{Andrea.Caviglia@unige.ch}
	\affiliation[University of Geneva]{Department of Quantum Matter Physics, University of Geneva, Geneva, Switzerland}

	\date{\today}

\begin{document}
\maketitle
\begin{abstract}
We investigate tunable superconducting transitions in (111)LaAlO$_3$/KTaO$_3$ field-effect devices. Large increases in conductivity, associated with superconducting fluctuations, are observed far above the transition temperature. However, the standard Aslamazov-Larkin paraconductivity model significantly underestimates the effect observed here. 
We use a model that includes conductivity corrections from normal state quasiparticle interference together with all contributions from superconducting fluctuations evaluated at arbitrary temperatures and in the short-wavelength limit.
Through analysis of the magnetoconductance and resistive transitions, we find that the large conductivity increase can be explained by a combination of weak anti-localization and Maki-Thompson superconducting fluctuations. Both contributions are enabled by a strong temperature dependence of the electron's decoherence time compatible with an electron-phonon scattering scenario. We find that conductivity corrections are modulated by the electrostatic field effect, that governs a competition between normal-state quasiparticle interference and superconducting fluctuations.

\end{abstract}

\section*{Introduction}

The discovery of a two-dimensional electron system at the LaAlO$_3$/SrTiO$_3$ interface \cite{ohtomo_high-mobility_2004} and its electrically controllable superconductivity~\cite{caviglia_electric_2008} sparked interest into emergent phenomena at oxide interfaces \cite{zubko_interface_2011}. Superconductivity in KTaO$_3$ was initially discovered in (001) oriented liquid ion gated devices \cite{ueno_discovery_2011}, and more recently at the interface with a plethora of oxides (LaAlO$_3$ \cite{liu_two-dimensional_2021,chen_electric_2021}, LaVO$_3$\cite{Liu2023}, EuO \cite{liu_two-dimensional_2021,liu_tunable_2023}, LaMnO$_3$ \cite{al-tawhid_enhanced_2023}, Al$_2$O$_3$ \cite{kumar_ojha_flux-flow_2023}). Unlike in SrTiO$_3$, superconductivity in KTaO$_3$ is highly orientation dependent \cite{liu_tunable_2023}, with a critical temperature ($T_\mathrm{c}$) of up to $\approx 2$~K for the (111) orientation, $\approx 1$~K for (110), and so far no transition has been observed for the (100) interface in solid state devices, or in the doped bulk material. Superconductivity at KTaO$_3$ interfaces arises from a combination of surface-induced symmetry breaking and chemical doping. However, electronic decoherence and fluctuations in superconducting field-effect devices remain largely unexplored.

A peculiar feature of superconductivity of (111) KTaO$_3$ interfaces is the occurrence of broad resistive transitions which, in our case, can be observed in a large range of temperatures far above $T_\mathrm{c}$ (see Figure~\ref{fig:2}a and \cite{liu_two-dimensional_2021,liu_tunable_2023, chen_electric_2021,mallik_superfluid_2022}). This broadening is usually referred to as an `excess conductivity' and ascribed to superconducting fluctuations above $T_\mathrm{c}$, most commonly within the Aslamazov-Larkin (AL) paraconductivity framework \cite{aslamasov_influence_1968}, however, this alone cannot account for the amount of excess conductivity observed in KTaO$_3$ interfaces. AL paraconductivity only accounts for the direct contribution of fluctuating Cooper pairs forming just above the critical temperature in the limit of long-wavelength and low-frequency fluctuations, therefore limiting its applicability to $T\gtrsim T_\mathrm{c}$ \cite{varlamov_fluctuation_2018}. Another common model adopted for the study of fluctuating superconductivity is the Halperin-Nelson formula \cite{halperin_resistive_1979}. This is an interpolation between the Berezinski-Kosterlitz-Thouless \cite{kosterlitz_ordering_1973} and simplified AL formulas, so faces similar limitations.

To solve this problem, we adopt the more refined framework of fluctuation spectroscopy \cite{glatz_fluctuation_2011}, in which indirect, short-wavelength and high-frequency (dynamical) fluctuations are taken into account. 
In addition to AL contribution~\cite{aslamasov_influence_1968}, the framework considers the anomalous Maki-Thompson (MT) process, which describes single-particle quantum interference in the presence of superconducting fluctuations~\cite{maki_critical_1968, thompson_microwave_1970}. The first order MT corrections to conductivity near $T_\mathrm{c}$ can be expressed through the analytical formula \cite{reizer_fluctuation_1992,thompson_microwave_1970}:
\begin{equation}
G^{MT, (0)}\propto \frac{2 \pi k_\mathrm{B} T}{\hbar\tau_\mathrm{GL}^{-1} - \hbar\tau_\phi^{-1}}\ln \left(\frac{\tau_\phi}{\tau_\mathrm{GL}}\right)
\end{equation}
where $\tau_\phi$ represents the electron's decoherence time and $\tau_\mathrm{GL}$ represents the Ginzburg-Landau dynamical time defined as $\hbar\tau_\mathrm{GL}^{-1} = k_\mathrm{B}T \ln\left(T/T_\mathrm{c}\right)$.
The fluctuation spectroscopy framework also includes changes in conductivity in the presence of superconducting fluctuations due to a depletion of the single-particle density of states (DOS) and diffusion coefficient renormalization (DCR)~\cite{ioffe_effect_1993, dorin_fluctuation_1993}. Finally, dealing with a two-dimensional electron system with significant electron scattering, we also include a corresponding correction for weak localization, $G^{\mathrm{WL}} \approx -\frac{e^2}{\pi h}\ln\left({\tau_\phi}/{\tau_\mathrm{tr}}\right)$, and weak anti-localization, $G^\mathrm{WAL} \approx \frac{e^2}{\pi h}\ln\left( \left( 1 + \tau_\phi/\tau_\mathrm{so} \right) \left( 1 + 2\tau_\phi/\tau_\mathrm{so}\right) ^{1/2} \right)$ \cite{bruus_many-body_2004}, which depends on both the elastic scattering time $\tau_\mathrm{tr}$, the phase-coherence time $\tau_\phi$, and the spin-orbit scattering time $\tau_\mathrm{so}$.
The full conductivity formula is then given by:
\begin{equation}
	G = G^{(0)} + G^{\mathrm{WL}} + G^{\mathrm{WAL}} + G^{\mathrm{AL}} + G^{\mathrm{MT}} + G^{\mathrm{DOS}} + G^{\mathrm{DCR}}
	\label{cond_exp}
\end{equation}
where $G^{(0)}$ incorporates other conductivity contributions whose temperature dependence can be neglected for $T<10$~K. 
To avoid over-fitting, we constrain the problem by estimating the scattering times $\tau_\mathrm{tr}$, $\tau_\phi$ and $\tau_\mathrm{so}$ through the analysis of the magnetoconductance at 10~K (where the contribution of superconducting fluctuations can be neglected). We then assume $\tau_{\mathrm{tr}}$ and $\tau_{\mathrm{so}}$ to be temperature independent, while the phase scattering time has a power-law temperature dependence: $\tau_\phi \propto T^\alpha$. Using these constraints, we fit the resistive transitions using three free parameters: $T_\mathrm{c}$, $R_0 = 1/G^{(0)}$ and $\alpha$. The fitting of experimental data with Eq.~(\ref{cond_exp}) allows us to extract important material parameters of the system and map out their behavior across the electronic phase diagram. 

Notably we find that, among all the superconducting contributions to the resistive transition, the MT component appears to dominate at all temperatures across the phase diagram. Moreover,in the region $T>2T_\mathrm{c}$, the superconducting fluctuations are small compared to the normal-state quasi-particle interference contributions (WL and WAL) which are responsible for the rounding of the resistive transitions far above $T_\mathrm{c}$.


\section*{Theoretical background}

In the standard Kubo formalism, the relationship between the electric current and the vector potential is established through the electromagnetic response operator: $\mathbf{j}=-\hat{Q}\mathbf{A}$. 
Graphically, within the framework of the diagrammatic technique at finite temperatures, the $\hat{Q}$ operator can be depicted by a loop diagram comprising two electron Green's functions connected through electromagnetic vertices.

Incorporating fluctuation pairing introduces a renormalization of both the Green's functions and the vertices by interactions in the Cooper channel (see Figure \ref{fig:1}a for the AL and MT components, the diagrams for all contributions are reported in the supplemental information \cite{supplementary}). Additionally, averaging over impurity positions has to be taken into account. This leads to ten leading-order corrections to the electromagnetic response operator, each containing a small parameter of the fluctuation theory (Ginzburg-Levanyuk number) as a prefactor.

The fluctuation correction to conductivity is determined by the imaginary part of the sum of all these contributions to the response operator, i.e.,
$G^\mathrm{fl}(T) = -\lim_{\omega \rightarrow 0}\Im Q^{\mathrm{(fl)}}(\omega, T)/\omega$.

The impact of superconducting fluctuations (SF) on conductivity near the superconducting critical temperature $T_{\mathrm{c}}$ is commonly analyzed in terms of three main contributions: the Aslamazov-Larkin (AL) process, the anomalous Maki-Thompson (MT) process, and the modification of the single-particle density of states (DOS) due to their involvement in the pairing of fluctuation Cooper pairs. The AL and MT processes lead to positive singular contributions to conductivity (diagrams 1 and 2 in Figure \ref{fig:1}a). In contrast, the DOS process depletes single-particle excitations at the Fermi level, resulting in a reduction of the Drude conductivity (diagrams 3--6\cite{supplementary}). The latter contribution is less singular, exhibiting only a logarithmic dependence on temperature close to $T_{\mathrm{c}}$, but becomes significant far from the critical temperature. Diagrams 7--10 \cite{supplementary} represent the renormalization of the diffusion coefficient (DCR diagrams) due to the presence of fluctuations, which are non-singular close to $T_{\mathrm{c}}$ but can be crucial far from the critical temperature.

The details of calculations for all temperatures and magnetic fields can be found in Refs.~\cite{glatz_fluctuation_2011,varlamov_fluctuation_2018}. In our case, when the magnetic field is zero, the final complete expression can be  written as

	\begin{eqnarray}
		G^\mathrm{fl}(t) & = & \underbrace{\frac{e^{2}}{\pi\hbar}\int\limits _{0}^{Y}ydy  \int\limits _{-\infty}^{\infty}\frac{dx}{\sinh^{2}\pi x}\frac{\left[\Re^{2}\mathcal{E}'-\Im^{2}\mathcal{E}'\right]\Im^{2}\mathcal{E}-2\Re\mathcal{E}'\Im\mathcal{E}'\Re\mathcal{E}\Im\mathcal{E}}{\left[\Re^{2}\mathcal{E}+\Im^{2}\mathcal{E}\right]^{2}}}_{G^\mathrm{AL}}\nonumber \\
		&  & +\underbrace{\frac{\pi e^{2}}{4\hbar}\int\limits _{0}^{Y}\frac{dy}{\delta+\frac{\pi^{2}}{2}y}\int\limits _{-\infty}^{\infty}\frac{dx}{\sinh^{2}\pi x}\frac{\Im^{2}\mathcal{E}}{\Re^{2}\mathcal{E}+\Im^{2}\mathcal{E}}}_{G^\mathrm{MT^{(an)}}}+\underbrace{\frac{e^{2}}{\pi^{2}\hbar}\int\limits _{0}^{Y}dy{\sum_{k=0}^{\infty}}'\frac{\mathcal{E}''(t,k,y)}{\mathcal{E}(t,k,y)}}_{G^\mathrm{MT^{(reg)}}+G^\mathrm{DOS^{(1)}}}\;\nonumber \\
		&  & +\underbrace{\frac{e^{2}}{\pi\hbar}\int\limits _{0}^{Y}dy\int\limits_{-\infty}^{\infty}\frac{dx}{\sinh^{2}\pi x}\frac{\Im\mathcal{E}\Im\mathcal{E}'}{\Re^{2}\mathcal{E}+\Im^{2}\mathcal{E}}}_{G^\mathrm{DOS^{(2)}}}+\underbrace{\frac{e^{2}}{6\pi^{2}\hbar}\int\limits _{0}^{Y}ydy{\sum_{k=0}^{\infty}}'\frac{\mathcal{E}'''(t,k,y)}{\mathcal{E}(t,k,y)}}_{G^\mathrm{DCR}}\,.\label{all0}
	\end{eqnarray}

with $\mathcal{E}\equiv\mathcal{E}(t,\imath x,y)=\ln t+\psi\left[\frac{1+\imath x}{2}+y\right]-\psi\left(\frac{1}{2}\right)$,
$\mathcal{E}^{(n)}(t,\imath x,y)\equiv\frac{\partial^{n}}{\partial y^{n}}\mathcal{E}(t,\imath x,y)=\psi^{(n)}\left[\frac{1+\imath x}{2}+y\right]$,
$\psi$ is the digamma function, the upper limit $Y=\hbar/\left(k_\mathrm{B}T\tau_\mathrm{tr}\right)=\hbar/\left(tk_\mathrm{B}T_\mathrm{c}\tau_\mathrm{tr}\right)$, the reduced temperature $t = T/T_\mathrm{c}$,
the phase-breaking rate $ \delta = \pi\hbar/(8k_\mathrm{B}T\tau_{\phi})$,
 and the prime-sum is taken with a factor $1/2$ for the $k=0$ summand. See supplemental information \cite{supplementary} for further information about this model as well as links to the code repositories.

\section*{Results and discussion}

\begin{figure}[tbh]
    \centering
    \includegraphics[width=\textwidth]{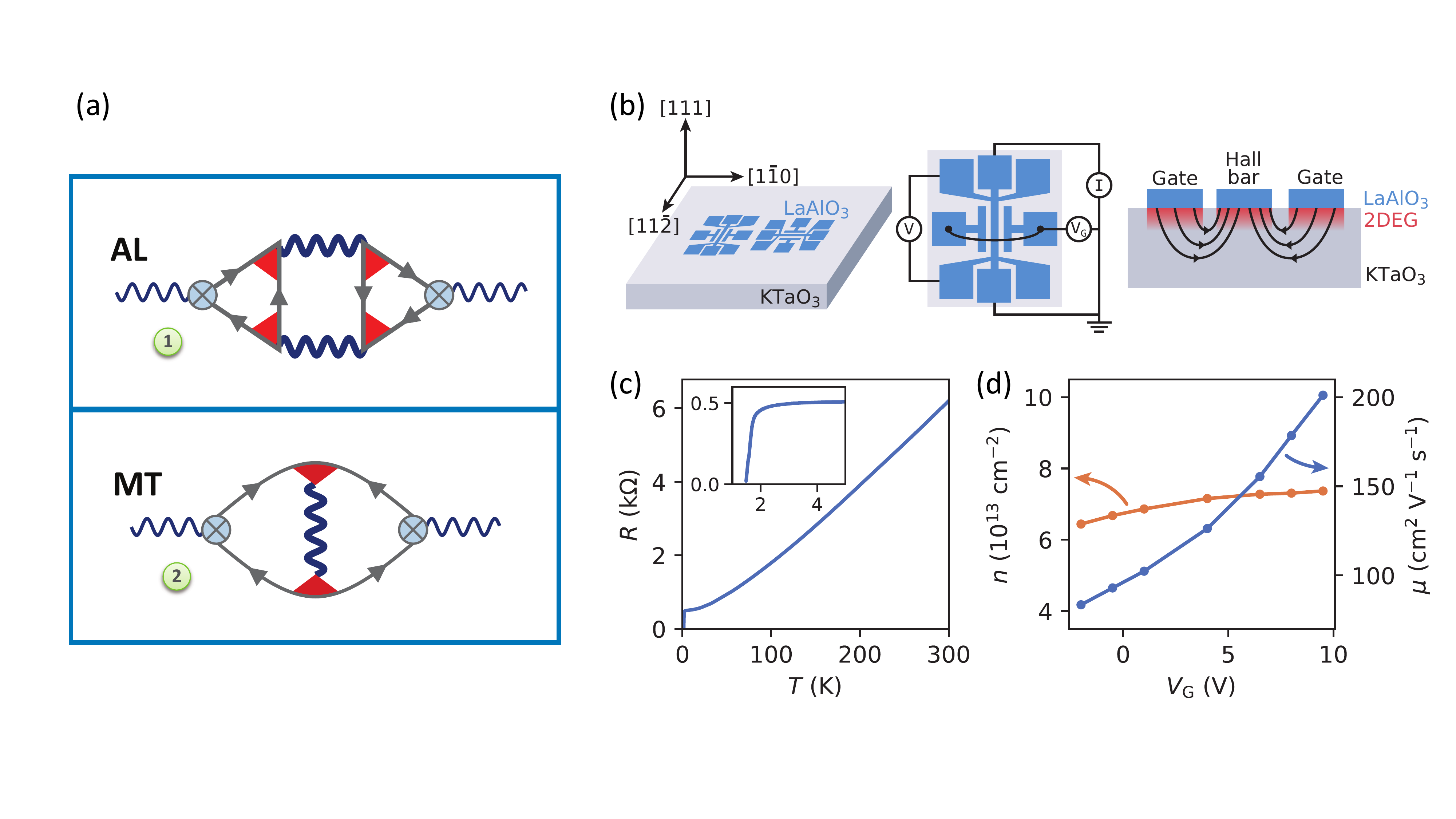}
    \caption{(a) Feynman diagrams for the leading-order contributions to Aslamazov-Larkin (1) and Maki-Thompson (2) contributions. Wavy lines correspond to fluctuation propagators, solid lines with arrows represent impurity-averaged normal state electron Green's functions, crossed circles are electric field vertices and triangles are impurity ladders accounting for the electron scattering at impurities (Cooperons) (From Ref.~\cite{glatz_fluctuation_2011}). (b) Device configuration, Hall bars are patterned on KTaO$_3$ (111) via electron-beam lithography and argon milling etching of the amorphous LaAlO$_3$ overlayer. A conducting 2D electron gas (2DEG) is preserved only below the interface between KTaO$_3$ and a-LaAlO$_3$. (c) Sheet resistance vs temperature for the virgin state 2DEG, inset zoomed on superconducting transition. (d) Carrier density $n$ and mobility $\mu$ versus back-gate at $T=10$~K. 
    Carrier density is found from Hall resistance measurements of $\pm12$~T and the mobility is calculated from $\mu=1/e R_\square n$. From $-2$ to $9.5$~V the carrier density remains roughly constant while the mobility more than doubles, so the large change in sheet resistance with gate is primarily due to the change in scattering rate and/or effective mass.}
    \label{fig:1}
\end{figure}

We fabricate Hall bar devices on KTaO$_3$ (111)-oriented substrates as illustrated in Fig.~\ref{fig:1}a, and measure sheet resistance vs temperature [$R_\square(T)$] and Hall effect at various gate voltages (see Methods \cite{supplementary}). In the main text we show data from a ``local back-gated'' device, however these results were reproduced also in a ``global back-gated'' device (see \cite{supplementary} for further discussion on the gating geometry and additional transport measurements).

\begin{figure*}
    \centering
    \includegraphics[width=\textwidth]{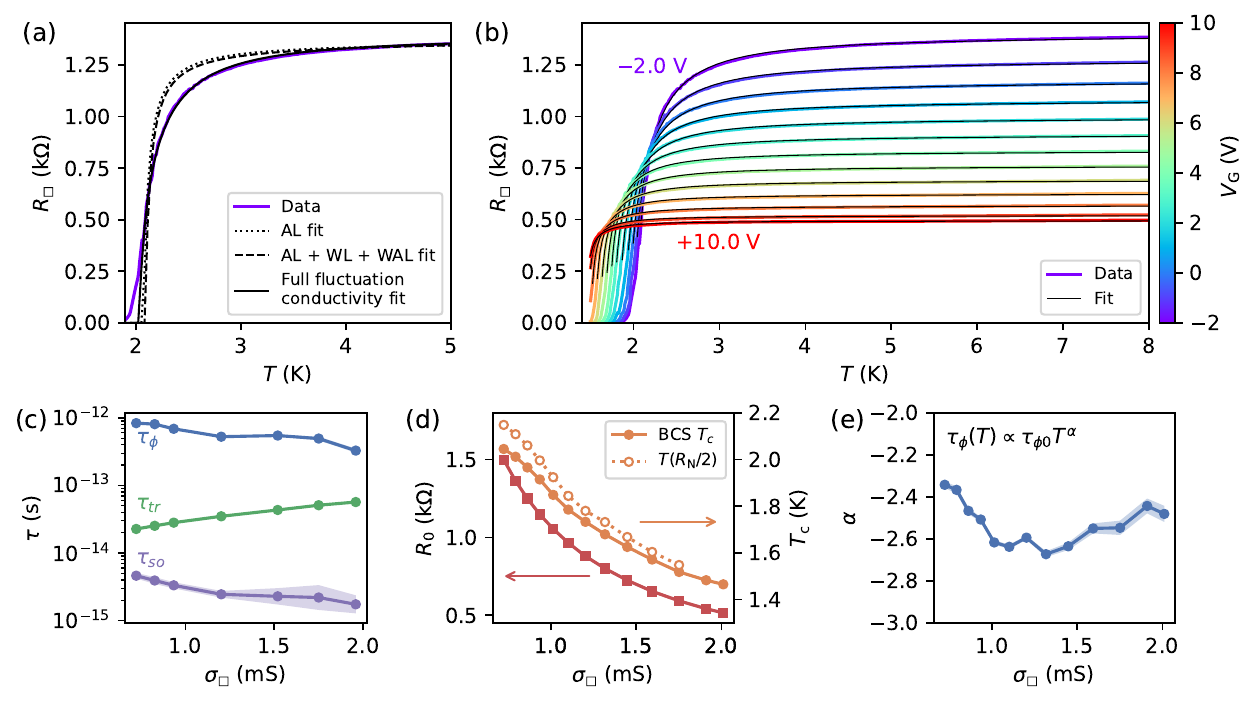}
    \caption{(a) Superconducting transition at $-2$~V, comparing the best fit from Aslamazov-Larkin and the full fluctuation conductivity model. (b) Superconducting transition at different gate voltages (colored lines), fit to the full fluctuation conductivity model (black lines). (c) Scattering times found from magnetoconductance analysis at 10~K, using $m^* \in [0.4m_e, 0.6m_e]$ (indicated by the shaded regions). These scattering times are used as constraints in the $R(T)$ fitting. The phase scattering time at 3K is computed using the value of $\alpha$ obtained from the fitting routine. (d) Fit results for the residual resistance $R_0$ and mean-field superconducting transition temperature $T_\mathrm{c}$. The more commonly adopted definition of critical temperature as the midway point of the resistive transition is also shown for comparison (dashed line).  (e) The temperature exponent, $\alpha$, for the phase breaking scattering time $\tau_\phi(T) = \tau_{\phi}(10\mathrm{K}) (T/10\mathrm{K})^{\alpha}$. $\tau_\phi(10~\mathrm{K})$ is fixed by magneto-conductance fitting. The shaded regions represent the range of output values due to the range of values in the input parameters.}
    \label{fig:2}
\end{figure*}

A metallic 2DEG forms at the interface between LaAlO$_3$/KTaO$_3$ with a superconducting transition near 2~K, as shown in Fig.~\ref{fig:1}b. These devices undergo a `gate-forming' process, where applying a positive back-gate voltage causes large irreversible changes in transport properties (which can be recovered by thermal cycling the sample to room temperature). This is likely due to the irreversible motion of charged defects such as oxygen vacancies.
After gate-forming, $R_\square$ and $T_\mathrm{c}$ at zero-gate irreversibly increases and the superconducting transition becomes more uniform. The transition is then reversibly gate-tunable and $T_\mathrm{c}$ is enhanced as the normal-state conductance is decreased, i.e. negative back-gate applied, as can be seen in Fig.~\ref{fig:2}b. Fig.~\ref{fig:1}c shows the effect of back-gate voltage on carrier density and mobility of the 2DEG, plotted on the same relative scale, showing that the change in sheet resistance is almost entirely caused by changes in mobility rather than carrier density. Similar effects have been ascribed to a negative back-gate voltage increasing the effective disorder by pushing the electron gas closer to the more defective interface region \cite{chen_electric_2021, Hwang_dualgate}. The gate voltage also affects the permittivity of KTaO$_3$, as discussed for LaAlO$_3$/SrTiO$_3$ \cite{gariglio_electron_2015, Bell_2009} interfaces.

As with similar devices measured in literature, we observe a very rounded superconducting transition, implying a large excess conductivity above $T_\mathrm{c}$. In Fig.~\ref{fig:2}a we analyze the resistive transition using the standard Aslamazov-Larkin model, as well as Aslamazov-Larkin together with WL and WAL, but find that these models significantly underestimate the excess conductivity near the transition. 

Before we apply the extended fluctuation theory, we need to constrain the model parameters as much as possible. 
Our samples show clear signs of WAL in the magnetoconductance at 10~K \cite{supplementary} which we analyze using the Hikami-Larkin-Nagaoka (HLN) model \cite{Hikami1980}. The results of the analysis are shown in Figure~\ref{fig:2}c. We assume an effective mass of 0.5~$m_e$, consistent with the range of masses for the light electron band reported in literature \cite{bareille_two-dimensional_2014,santander-syro_orbital_2012,wang_prediction_2019,zhang_strain-driven_2018,mallik_electronic_2023, Varotto2022}. Different choices of effective mass systematically shift all scattering times and ultimately the temperature exponent by a similar relative amount. We are aware that the HLN model treats spin orbit scattering in a perturbative fashion, and that the results of our analysis go beyond this regime. However, we note that recent spectroscopic measurements for the Rashba splitting in (001) KTaO$_3$ reported $\alpha_\mathrm{R}\approx 320$~meV~\AA$^{-1}$ at $k_\mathrm{F}\approx 0.12$~\AA$^{-1}$ \cite{Varotto2022} corresponding to $\tau_\mathrm{so}=8.6$~fs which is comparable with the values that we obtain through magnetotransport analysis.

While we assume that $\tau_\mathrm{tr}$ and $\tau_{\mathrm{so}}$ are temperature independent, the same cannot be done for the electron's dephasing time $\tau_\phi$. There are a number of mechanisms that can induce electron dephasing, each with its own temperature dependence. Some of the most commonly treated electron dephasing sources are Nyquist dephasing ($\tau_\phi\propto T^{-1}$)\cite{Altshuller_81_nyquistoise}, electron-electron scattering from Coulomb interaction ($\tau_\phi\propto T^{-1}$) \cite{Abrahams1981, altshuler_quantum_1982, FukuyamaAbrahams_83, Fukuyama_84} and electron-phonon scattering ($\tau_\phi\propto T^{-2}$ or $\tau_\phi\propto T^{-3}$) \cite{Lawrence1978, Raffy1985, Giannouri1997}.
Finally, our resistive transitions show a 'foot' in the lower resistance half of the transition see Figure \ref{fig:2}a,b. This feature has been ascribed to inhomogeneities or BKT-like fluctuations \cite{Benfatto2009a,Caprara2011b, Maccari2017} that our model cannot capture. To mitigate the interference from these contributions, we limit the fit to $R_\square(T)>R_\square(T=10~\mathrm{K})/3$.
The fit parameters we vary are then $R_0 = 1/G^{(0)}$, the superconducting transition temperature $T_\mathrm{c}$ (Fig.~\ref{fig:2}d), and the temperature dependence of the phase-coherence time $\alpha$ where $\tau_\phi(T) \propto T^\alpha$ (Fig.~\ref{fig:2}e).
The extended fluctuation theory (see Eq. (\ref{cond_exp}),(\ref{all0}),\cite{supplementary}) yields a much better fit when compared with the AL model or AL+WL+WAL models showcasing the importance of the other superconducting fluctuations contributions (see Figure \ref{fig:2}a). The results of the full fluctuation analysis at every gate voltage is shown in Fig.~\ref{fig:2}b. 

From our analysis, we find, within error, a constant value of $\alpha \approx -2.5$ for the temperature dependence of the phase-coherence time (and $\alpha \approx -3$ in another sample, see \cite{supplementary}). A range of $\alpha$ has been observed in other systems \cite{Bergmann1984, Raffy1985, Giannouri1997, Lomakin2022} with values close to the one that we report. The enhanced temperature exponent $\alpha$ together with the gate dependence of $\tau_\phi(10\mathrm{K})$ render e-e interactions an unlikely cause of electron's decoherence. The most adopted model for e-e interaction induced dephasing predicts that $\tau_\phi \propto \frac{\sigma_\square}{\sigma_0}/\ln(\frac{\sigma_\square}{\sigma_0})$ ($\sigma_0 = e^2/\pi h$) \cite{altshuler_quantum_1982, FukuyamaAbrahams_83} meaning that a more conducting sample's state  corresponds with a longer coherence time; here we observe the opposite trend (Figure \ref{fig:2}c).

On the other hand, electron-phonon scattering (e-ph) from acoustic phonons has been predicted to cause electron decoherence with an enhanced temperature exponent \cite{Lawrence1978}. For the dirty 2D limit $\tau_\phi^{-1} \propto \lambda \omega_\mathrm{D}\big(\frac{T}{\Theta_\mathrm{D}}\big)^3$\cite{Raffy1985} where $\lambda$ is the electron-phonon coupling, $\omega_\mathrm{D}$ and $\Theta_\mathrm{D}$ are Debye's frequency and temperature respectively.
While the e-ph scattering $\tau_\phi$ is expected to be independent of $\sigma_\square$ or $n_{\mathrm{2D}}$, the gate dependence of $\tau_\phi$ can be associated with the electric field dependence of the dielectric permittivity of quantum paraelectrics like KTaO$_3$ \cite{Christen1994,Ang2001,Skoromets2016}. Studies on the quantum paraelectric SrTiO$_3$ show a decrease of dielectric constant with applied field, which enhances the electron-phonon coupling strength $\lambda$ while suppressing $\tau_\phi$ \cite{stornaiuolo_weak_2014, neville_permittivity_1972, matthey_field-effect_2003}.

This kind of interfaces have been proposed to host an unconventional e-ph coupling that goes beyond the standard one mediated by acoustic phonons. 
Recent theoretical developments have shown that the lowest transverse optical phonon TO1, associated with the quantum paraelectric behavior of KTaO$_3$, can also mediate a finite e-ph coupling provided that a strong symmetry breaking occurs \cite{Gastiasoro2022a}. These theories were quite successful in understanding the strong anisotropy in the superconducting pairing at the different interfaces of KTaO$_3$ \cite{liu_tunable_2023}.
Discerning which phonon is dominating the electron decoherence in KTaO$_3$ 2DEGs would require further theoretical and experimental investigations and goes beyond the scope of this work.
We conclude that a gate-tunable electron-phonon scattering is the main source of electron's decoherence in our experiments, independently of the particular phonon(s) involved in this process. 

Another possible cause of broad resistive transitions is an inhomogeneous superconducting layer. We model an inhomogeneous superconductor by simulating a Random Resistor Network (RRN) in which every node has a resistive transitions with a random $T_\mathrm{c}$ and $R_0$ but $\alpha = 1$ (see \cite{supplementary}). In accordance with other reports, we find that the resistive transitions of the whole RRN broaden and develop a `foot' at low temperatures\cite{Benfatto2009a,Caprara2011b, Maccari2017}, but they cannot be misunderstood for a transition of a uniform superconductor with the same scattering times and enhanced temperature exponent.

\begin{figure}[t]
    \centering
    \includegraphics[width=\textwidth]{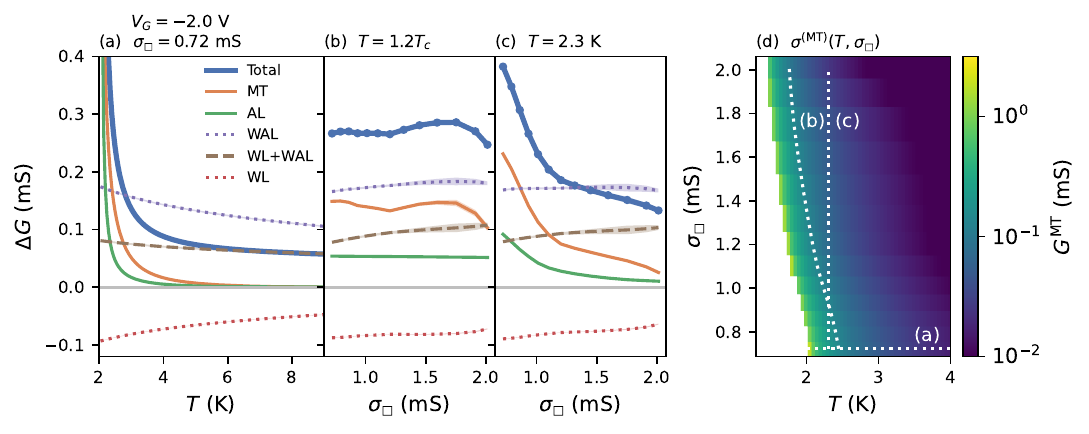}
    \caption{(a) Comparison of different contributions conductivity versus temperature at $-2$~V assuming $\tau_\phi = \tau_{\phi0} T^{-2.5}$. Here the effective mass of the electron has been fixed to $m^* = 0.5 m_\mathrm{e}$, the shaded colorband is caused only by the uncertainty in the determination of the scattering times. Due to the long phase coherence time, Maki-Thompson and weak-localization corrections are dominant over the whole temperature range, and Maki-Thompson corrections dominate the temperature dependence at lower temperatures. (b-c) Gate voltage control of the different contributions at (b) $T = 1.2 T_\mathrm{c}$ and (c) $T = 2.3$~K, showing the dominance of Maki-Thompson and weak localization over the whole gate range. (d) Map of Maki-Thompson contributions versus temperature and sheet conductance, white dashed lines indicate the plots in (a), (b), (c).}
    \label{fig:3}
\end{figure}

Having established that $\alpha$ is approximately constant throughout the gating range and that it is compatible with an e-ph scattering scenario, we fix $\alpha = -2.5$ and repeat the fitting of the resistive transitions. This step allows us to study the gate dependence of the contributions from superconducting fluctuations without convoluting them with the gate dependence of $\alpha$. The fit to the resistive transitions remains good, with only small modifications to the gate dependence of $R_0$ and $T_\mathrm{c}$ (see \cite{supplementary}).

The full fluctuation model allows us to disentangle the conductivity contributions both as a function of temperature and gate voltage (Figure \ref{fig:3}).
When considering the temperature dependence at fixed gate voltage (Figure \ref{fig:3}a) we can identify two separate regimes. For temperatures above $\approx3T_\mathrm{c}$ the excess conductivity is dominated by normal-state quasiparticle interference. Here $\tau_\phi \gg \tau_\mathrm{tr} \gg \tau_\mathrm{so}$ rendering localization contributions quite significant with WAL dominating due to the extremely short spin-orbit time ($\Delta \sigma \approx 80~\mu$S at 10~K).
For temperatures below $\approx2.5T_\mathrm{c}$, superconducting fluctuations dominate the temperature dependence with the Maki-Thompson fluctuations being the most relevant of the mechanisms considered (for clarity DCR and DOS are not shown here, see \cite{supplementary} for a full visualization of the parameters).
We observe a gate-tuned competition between the normal state quasi-particle interference and superconducting fluctuations contributions (Figure \ref{fig:3}b,c). At a fixed temperature, the change in fluctuation conductivity is convoluted with the gate induced change in critical temperature. However, by taking a slice at fixed reduced temperature ($T = 1.2T_\mathrm{c}$ for example) (Figure~\ref{fig:3}b), we observe a non-monotonous tuning of the Maki-Thompson contributions with a dip near 1.2~mS that is directly related to features in the gate dependence of $\tau_\phi$. 
Finally we note that, while $\tau_{so}$ appears to be the most prominent scattering rate in our samples, it contributes only to the WAL corrections and does not affect the superconducting fluctuations. The effects of the D'yakonov-Perel spin relaxation mechanism on superconducting fluctuations still lacks theoretical investigation, in particular in the presence of Rashba spin-orbit coupling in two dimensions.

\section*{Conclusion}

We observe highly gate-tunable superconductivity KTaO$_3$-based devices. Consistently with previous literature reports, we find a large excess conductivity which cannot be accounted for by the standard Aslamazov-Larkin model. Using magnetoconductance analysis and fitting our resistance vs temperature to a model considering short-wavelength and dynamical superconducting fluctuations, we find that this excess conductivity is primarily due to weak anti-localization at high temperatures, and Maki-Thompson corrections near the critical temperature.

Assuming a temperature dependence of the phase coherence time $\tau_\phi\propto T^\alpha$ and an effective mass 0.5$m_e$, we find an enhanced temperature exponent of $\alpha = -2.5 \pm 0.3$ which is almost constant throughout the studied gating range. This observation rules out electron-electron scattering as the dominant mechanism affecting electron's coherence in KTaO$_3$, and points at electron-phonon scattering as the most significant.

We find that the corrections to conductivity can be tuned with gate by means of a competition between normal state quasi-particle interference and superconducting fluctuations.
We showcase the tuning of the electron's dephasing time through the gate control of electron-phonon scattering and link it directly to the magnitude and tunability of superconducting fluctuations above the critical temperature. 
These results are an important step in understanding the phenomenology of two-dimensional superconductivity quantum paraelectric superconductors like KTaO$_3$.

\section{Acknowledgments}
This work was supported by the Swiss State Secretariat for Education, Research and Innovation (SERI) under contract no. MB22.00071, by the Gordon and Betty Moore Foundation (grant no. 332 GBMF10451 to A.D.C.), by the European Research Council (ERC), by the research program Open Competition ENW Groot financed by the Dutch Research Council (NWO, TOPCORE project number OCENW.GROOT.2019.048).
M.G. acknowledges funding from the Agence Nationale de la Recherche  (ANR project “SURIKAT”  ANR-23-CE30-0036).
A.G. was supported by the U.S. Department of Energy, Office of Science, Basic Energy Sciences, Materials Sciences and Engineering Division.

\paragraph{Data and Code Availability}
The data and code used to produce the figures for the main text and supplemental material are available in the \href{https://doi.org/10.5281/zenodo.15705548}{Zenodo Repository}.

\pagebreak

\bibliography{refs.bib,supplementary.bib,uldorefs.bib}
	
\end{document}


\maketitle
\newpage
\FloatBarrier
\section{Experimental methods}
\FloatBarrier

\textbf{Device Fabrication.} Conducting samples are prepared by pulsed laser deposition of 10~nm amorphous LaAlO$_3$ on (111)-oriented single-crystal KTaO$_3$ substrates. We use a pulsed KrF laser (248~nm) at 1~Hz with a fluence $\sim1$~J~cm$^{-2}$. The substrate temperature during deposition is maintained at 530~K and in a low oxygen atmosphere with a pressure of $10^{-4}$~Pa. After deposition, Hall bar devices are patterned on LaAlO$_3$/KTaO$_3$ films using e-beam lithography followed by argon milling of the amorphous LaAlO$_3$ overlayer. We measured comparable results in a number of devices, in the main text we show results from 3$\times$12~\textmu m Hall bar.

\textbf{Electrical Transport Measurements.} The sample is attached to a metallic electrode with a thin and uniform layer of silver paint, and ohmic contacts to the 2DEG were made using ultra-sonic aluminium wedge-bonding. The sample was cooled in He-flow cryostat with a base temperature of 1.5~K. Some samples show a hysteretic behavior in the $R_\square$ vs $V_\mathrm{G}$. When this occurred, in order to keep gating results consistent, all the measurements were performed on the up-sweep (going from negative to positive gate-voltage) never reversing the sweeping direction unless the two extreme voltages were reached.

\section{Full fluctuation conductivity model}

\begin{figure}[b!]
    \centering
    \includegraphics[width=0.8\columnwidth]{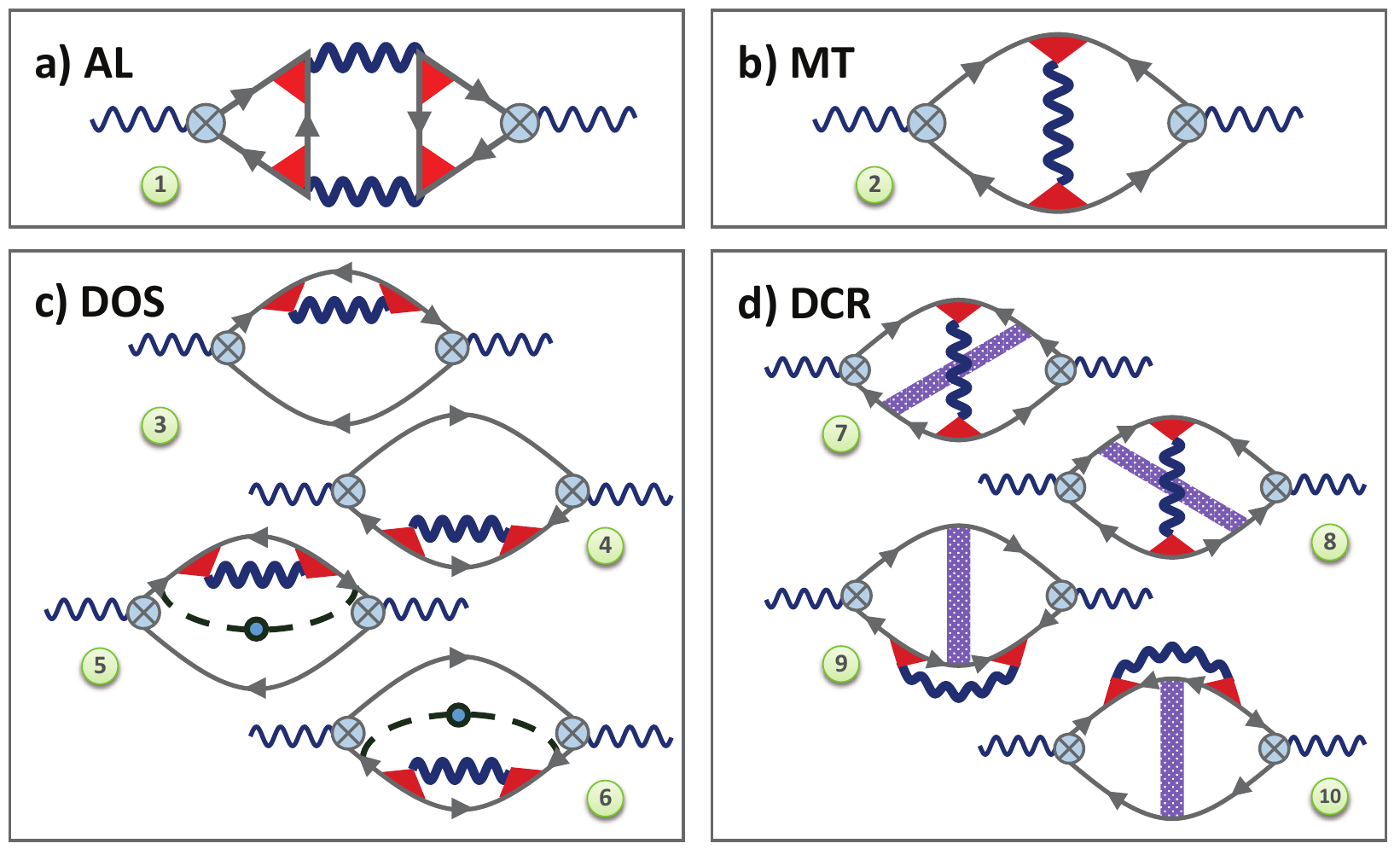}
    \caption{Feynman diagrams for the leading-order contributions to the electromagnetic response operator. Wavy lines correspond to fluctuation propagators, solid lines with arrows represent impurity-averaged normal state electron Green's functions, crossed circles are electric field vertices, dashed lines with a circle represent additional impurity renormalizations, and triangles and dotted rectangles are impurity ladders accounting for the electron scattering at impurities (Cooperons) (From Ref.~\cite{glatz_fluctuation_2011}).}
    \label{fig.conddia}
\end{figure}

A full discussion of the model can be found in Ref.~
\cite{glatz_fluctuation_2011}, relevant details are reproduced here.

In the standard Kubo formalism, the relationship between the electric current and the vector potential is established through the electromagnetic response operator:
\begin{equation}
    j_{\alpha} = -\int Q_{\alpha \gamma}(\mathbf{r, r^{\prime }}, t, t^{\prime})\mathbf{A}_{\gamma}( \mathbf{r^{\prime}}, t^{\prime})d\mathbf{r^{\prime }}dt^{\prime }\,.
    \label{kubo}
\end{equation}

Graphically, within the framework of the diagrammatic technique at finite temperatures, this operator is depicted by a loop diagram comprising two electron Green's functions connected through electromagnetic vertices.

Incorporating fluctuation pairing introduces a renormalization of both the Green's functions and the vertices by interactions in the Cooper channel, as illustrated by wavy lines in Figure~\ref{fig.conddia}. Additionally, averaging over impurity positions has to be taken into account. This leads to ten leading-order corrections to the electromagnetic response operator, each containing a small parameter of the fluctuation theory (Ginzburg-Levanyuk number) as a prefactor.

The fluctuation correction to conductivity is determined by the imaginary part of the sum of all these contributions to the response operator:
\begin{equation}
    \sigma^{\mathrm{(fl)}}(T) = -\lim_{\omega \rightarrow 0}\frac{\Im Q^{\mathrm{(fl)}}(\omega, T)}{\omega}\,.  \label{quom}
\end{equation}

The impact of superconducting fluctuations (SF) on conductivity near the superconducting critical temperature $T_{\mathrm{c}}$ is commonly analyzed in terms of three main contributions: the Aslamazov-Larkin (AL) process, the anomalous Maki-Thompson (MT) process, and the modification of the single-particle density of states (DOS) due to their involvement in the pairing of fluctuation cooper pairs (FCPs). The AL and MT processes lead to positive singular contributions to conductivity (diagrams 1 and 2 in Figure~\ref{fig.conddia}). In contrast, the DOS process depletes single-particle excitations at the Fermi level, resulting in a reduction of the Drude conductivity (diagrams 3--6 in Figure~\ref{fig.conddia}). The latter contribution is less singular, exhibiting only a logarithmic dependence on temperature close to $T_{\mathrm{c}}$, but becomes significant far from the critical temperature. Diagrams 7--10 represent the renormalization of the diffusion coefficient (DCR diagrams) due to the presence of fluctuations, which are nonsingular close to $T_{\mathrm{c}}$ but can be crucial far from the critical temperature.

The details of calculations for all temperatures and magnetic fields can be found in Refs. \cite{glatz_fluctuation_2011,varlamov_fluctuation_2018}.
In our case, when the magnetic field is zero, the final complete expression can be  written as
	\begin{eqnarray}
		G^\mathrm{fl}(t) & = & \underbrace{\frac{e^{2}}{\pi\hbar}\int\limits _{0}^{Y}ydy  \int\limits _{-\infty}^{\infty}\frac{dx}{\sinh^{2}\pi x}\frac{\left[\Re^{2}\mathcal{E}'-\Im^{2}\mathcal{E}'\right]\Im^{2}\mathcal{E}-2\Re\mathcal{E}'\Im\mathcal{E}'\Re\mathcal{E}\Im\mathcal{E}}{\left[\Re^{2}\mathcal{E}+\Im^{2}\mathcal{E}\right]^{2}}}_{G^\mathrm{AL}}\nonumber \\
		&  & +\underbrace{\frac{\pi e^{2}}{4\hbar}\int\limits _{0}^{Y}\frac{dy}{\delta+\frac{\pi^{2}}{2}y}\int\limits _{-\infty}^{\infty}\frac{dx}{\sinh^{2}\pi x}\frac{\Im^{2}\mathcal{E}}{\Re^{2}\mathcal{E}+\Im^{2}\mathcal{E}}}_{G^\mathrm{MT^{(an)}}}+\underbrace{\frac{e^{2}}{\pi^{2}\hbar}\int\limits _{0}^{Y}dy{\sum_{k=0}^{\infty}}'\frac{\mathcal{E}''(t,k,y)}{\mathcal{E}(t,k,y)}}_{G^\mathrm{MT^{(reg)}}+G^\mathrm{DOS(1)}}\;\nonumber \\
		&  & +\underbrace{\frac{e^{2}}{\pi\hbar}\int\limits _{0}^{Y}dy\int\limits_{-\infty}^{\infty}\frac{dx}{\sinh^{2}\pi x}\frac{\Im\mathcal{E}\Im\mathcal{E}'}{\Re^{2}\mathcal{E}+\Im^{2}\mathcal{E}}}_{G^\mathrm{DOS(2)}}+\underbrace{\frac{e^{2}}{6\pi^{2}\hbar}\int\limits _{0}^{Y}ydy{\sum_{k=0}^{\infty}}'\frac{\mathcal{E}'''(t,k,y)}{\mathcal{E}(t,k,y)}}_{G^\mathrm{DCR}}\,.\label{all0}
	\end{eqnarray}

with $\mathcal{E}\equiv\mathcal{E}(t,\imath x,y)=\ln t+\psi\left[\frac{1+\imath x}{2}+y\right]-\psi\left(\frac{1}{2}\right)$,
$\mathcal{E}^{(n)}(t,\imath x,y)\equiv\frac{\partial^{n}}{\partial y^{n}}\mathcal{E}(t,\imath x,y)=\psi^{(n)}\left[\frac{1+\imath x}{2}+y\right]$,
the upper limit $Y=1/\left(T\tau\right)=1/\left(tT_{c}\tau\right)$, the reduced temperature $t = T/T_{c}$,
the phase-breaking rate $ \delta = \pi\hbar/(8k_\mathrm{B}T\tau_{\phi})= \pi\hbar/(8tk_\mathrm{B}T_\mathrm{c}\tau_{\phi})$,  and the prime-sum taken with factor $1/2$ for the $k=0$ summand.

The different contributions to the conductivity are calculated numerically based on the FSCOPE C++ program available at \url{https://github.com/andreasglatz/FSCOPE}. An optimized Rust port and Python wrapper of these functions are used to perform a least-squares fit to the data. This Python library is available on PyPI, and can be installed with `\texttt{pip install fluctuoscopy}', the source code and documentation are available at \url{https://github.com/g-kimbell/fluctuoscopy}.

\FloatBarrier
\section{Additional Electrical Transport Data and Analysis}

\subsection{Magnetoconductance Analysis}
\FloatBarrier

\begin{figure}[ht]
    \centering
    \includegraphics[width = 15cm]{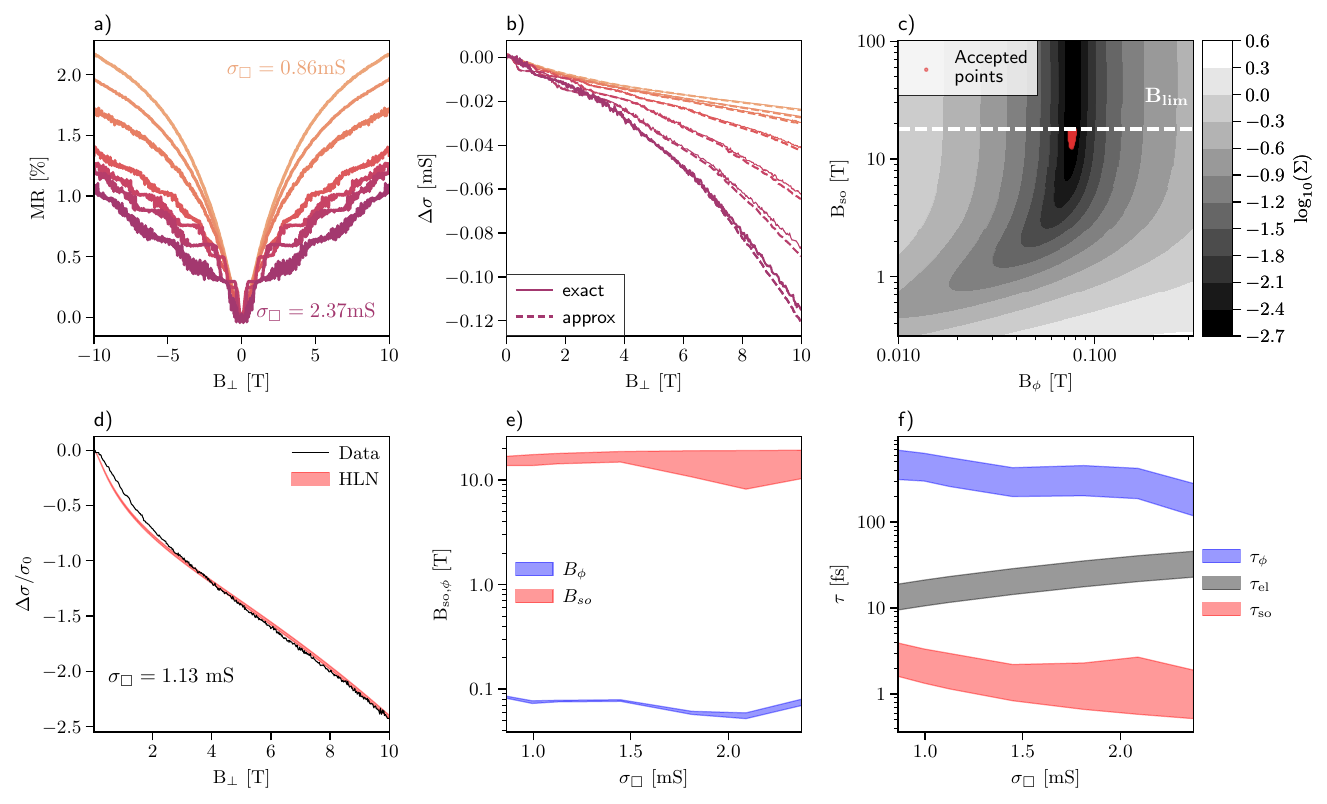}
    \caption{\textbf{Magnetoconductance analysis. (a)} Magnetoresistance traces for several gate voltage values. \textbf{(b)} Comparison between exact expression for $\sigma_{xx}$ (solid line) and the approximation described in equation~\ref{eq:sigma_approx} (dashed line). \textbf{(c)} Color map of the error function $\Sigma$. Overlaid in red, the range of parameters accepted as optimal. The white dashed line represents $B_{\mathrm{lim}}$. \textbf{(d)} Magnetoconductance data overlaid with the HLN model computed for all the values within the accepted parameters' range. \textbf{(e)} Plot of the effective fields as a function of sheet conductance. \textbf{(f)} Scattering times as a function of sheet conductance.}
    \label{fig:Supp2}
\end{figure}

In the following we describe the fitting procedure analogue to the one outlined in \cite{Filippozzi2024} that we had to adopt to overcome challenges in the fitting of KTaO$_3$(111) magneto-conductance.
All measurements of magnetoconductance are made at $T=10$~K, where we can neglect magnetic-field-dependent conductivity contributions from superconducting fluctuations.
In measurements at 10~K, the samples show clear signs of weak anti-localization (see Figure~\ref{fig:Supp2}a) which is a hallmark the effects of spin orbit scattering on the transport properties of electrons.
In order to obtain an estimate of the strength of spin orbit scattering in our sample, we analyze the magnetoconductance traces using the Hikami-Larkin-Nagaoka (HLN) \cite{Hikami1980} model adapted for the D'yaknov-Perel spin relaxation mechanisms:
\begin{equation}
\frac{\Delta\sigma(B)}{\sigma_0} = -\frac{1}{2}\Psi\bigg(\frac{B_\mathrm{\phi}}{B}\bigg)+\Psi\bigg(\frac{B_\phi+B_\mathrm{so}}{B}\bigg)+\frac{1}{2}\Psi\bigg(\frac{B_\phi+2B_\mathrm{so}}{B}\bigg)
\end{equation}
where $\Psi(x) = \psi(1/2 + x) -\log(1/x)$.
While a quadratic term is clearly visible in our magnetoconductance traces (see Figure \ref{fig:Supp2}b), we are able to estimate the magnitude of the quadratic component by means of the following approximation:
\begin{equation}
\label{eq:sigma_approx}
\sigma_\mathrm{xx}(B) = \frac{R_\mathrm{xx}(B)}{R_\mathrm{xx}^{2}(B)+R_\mathrm{xy}^2(B)} \approx \frac{1}{R_\mathrm{xx}(B)} - \frac{R_\mathrm{H}^2}{R_\mathrm{xx}^3(0)}B^2
\end{equation}
Figure~\ref{fig:Supp2}b also confirms the validity of this approximation given the good agreement between the dashed and solid lines. The amplitude of the quadratic contribution is fixed to $R_\mathrm{H}^2/R_\mathrm{xx}(0)^3$.

In Figure \ref{fig:Supp2}c we plot the error function $\Sigma$ as a function of the two free parameters $B_{so}$ and $B_{\phi}$ to quantify the agreement between the model and our data:
\begin{equation}
\Sigma(B_\mathrm{so}, B_\phi) = \frac{1}{N}\sum_{B_k} (\sigma(B_k)-\sigma_{\mathrm{HLN}}(B_k))^2
\end{equation} 
This plot shows how a simple least squares routing is inadequate to find the optimal parameters.

We define the optimal parameters as a set of $(B_\mathrm{so}, B_\phi)$ that follows the rule:
\begin{equation}
\Sigma(B_\mathrm{so}, B_\phi)< t \quad \mathrm{and}\quad B_\mathrm{so}<B_\mathrm{lim}
\end{equation}
where $t$ is a threshold proportional to the minimum error observed and  $B_\mathrm{lim}$ is defined with a very loose limit, namely $l_\mathrm{so}>\lambda_\mathrm{F}$:
\begin{equation}
\sqrt{\frac{\hbar}{4eB_\mathrm{so}}}>\lambda_\mathrm{F} \quad  \Rightarrow \quad
B_\mathrm{so}<\frac{\hbar}{4e\lambda_{F}^{2}}
\end{equation}
where $\lambda_\mathrm{F}$ represents the Fermi wavelength.
The resulting range of optimal parameters is shown in Figure~\ref{fig:Supp2}e. 
The characteristic scattering fields can be converted into scattering times:
\begin{equation}
\tau_\mathrm{so, \phi} = \frac{\hbar}{4D e B_\mathrm{so,\phi}}
\end{equation}
where $D = v_\mathrm{F}^{2}\tau_\mathrm{tr}/2$ and $\tau_\mathrm{tr}$ represents the bare transport time which enters the expression for the zero field conductivity as:
\begin{equation}
\sigma(0) = \frac{n_{\mathrm{2D}}e^2\tau_\mathrm{tr}}{m^*}-\sigma_0\log\bigg(\frac{\tau_\phi}{\tau_\mathrm{tr}}\bigg)+\sigma_0\log\bigg[\bigg(1+\frac{\tau_\phi}{\tau_\mathrm{so}}\bigg)\bigg(1+\frac{2\tau_{\phi}}{\tau_\mathrm{so}}\bigg)^{1/2}\bigg]
\end{equation}
The range of scattering times displayed in Figure \ref{fig:Supp2}f accounts for the uncertainty in the choice of optimal fitting parameters ($B_\mathrm{so}, B_\phi$) but also for a choice of electron's effective mass $m^*\in[0.2, 0.4]m_\mathrm{e}$ that accounts for the different values concerning the light electron band that are reported in literature. 

\FloatBarrier
\subsection{More fluctuation conductivity details}
\FloatBarrier

\begin{figure}[ht]
    \centering
    \includegraphics[width=\linewidth]{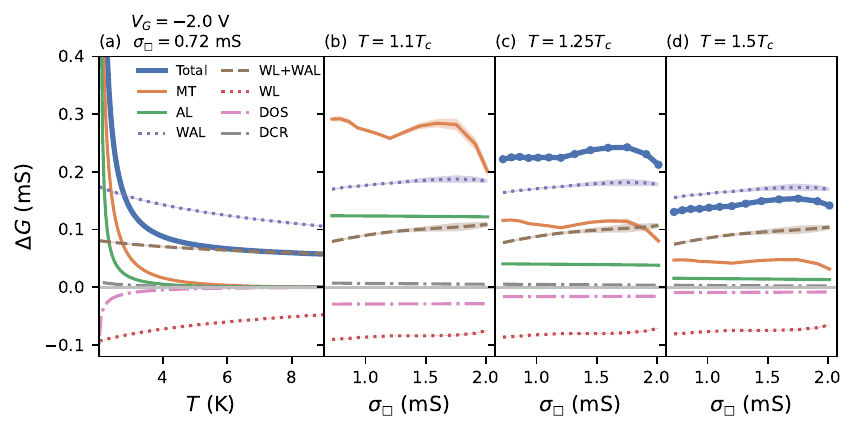}
    \caption{\textbf{All excess conductivity contributions} \textbf{(a)} as a function of temperature at $-2$~V gate and \textbf{(b)}-\textbf{(e)} as a function of normal state sheet conductance (controlled by gate voltage) at different reduced temperatures. All of these plots are computed assuming $\tau_\phi \propto T^{-2.5}$}
    \label{fig:KT027_fluc_components}
\end{figure}

Figure~\ref{fig:KT027_fluc_components} shows all the conductivity components from the model, including diffusion coefficient renormalization (DCR) and density of states (DOS) corrections, at different reduced temperatures. DCR and DOS are not shown in the main text.

\begin{figure}
    \centering
    \includegraphics[width=\linewidth]{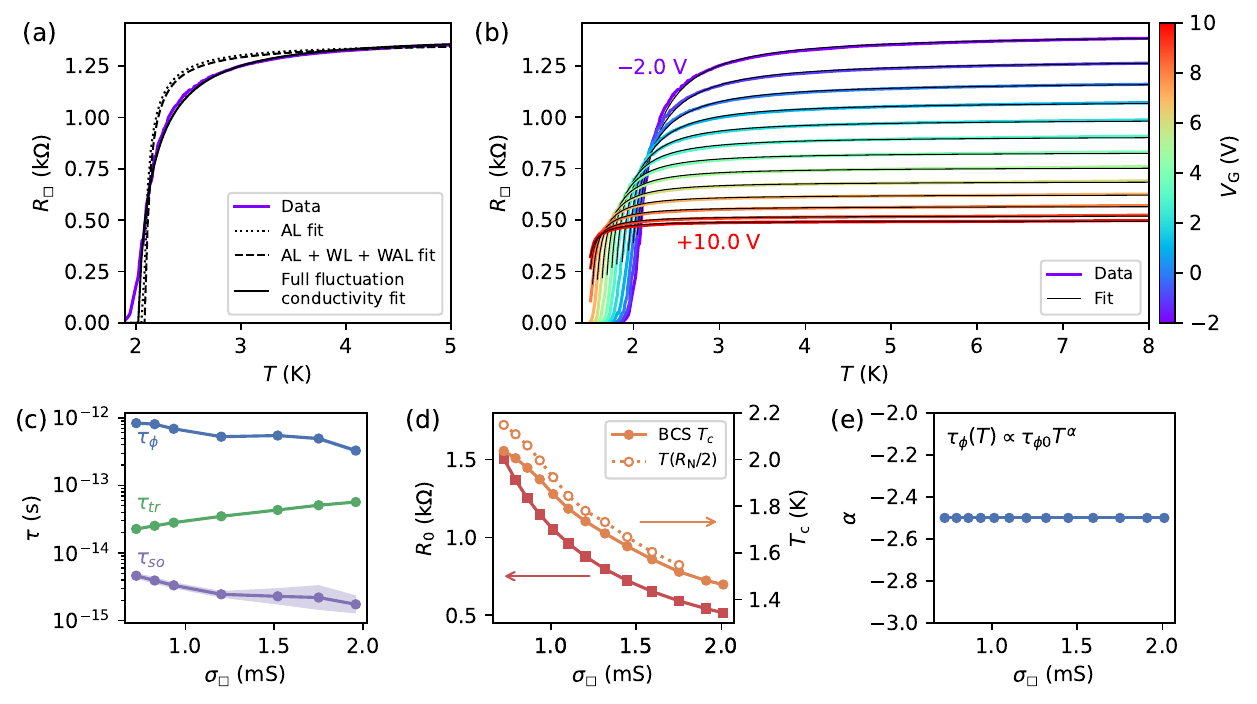}
    \caption{\textbf{Fitting of the resistive transitions at fixed alpha.} Here we repeat the analysis of the resistive transition as outlined in the text but we fix $\alpha = -2.5$. This choice does not significantly affect the results of the fitting. \textbf{(a)} The full fluctuation model still outperforms the other models considered.\textbf{(b)} The analysis of the resistive transitions works well throughout the whole gating range. \textbf{(c)} Scattering times, identical to the main text. \textbf{(d)} we see that the transition temperature and $R_0$ follows the same trend and show roughly the same values. \textbf{(e)} $\alpha$ is fixed for these fits.}
    \label{fig:KT027-fixed-alpha}
\end{figure}

Figure~\ref{fig:KT027-fixed-alpha} shows the $R(T)$ fit results when fixing $\alpha=-2.5$, so only varying the parameters $T_\mathrm{c}$ and $R_0$, as opposed to the free $\alpha$ fit shown in the main text. The model still fits well to the data assuming fixed $\alpha$.

\FloatBarrier
\section{Measurement and analysis on globally gated sample}
\FloatBarrier

\begin{figure}[t]
    \centering
    \includegraphics[width=0.5\linewidth]{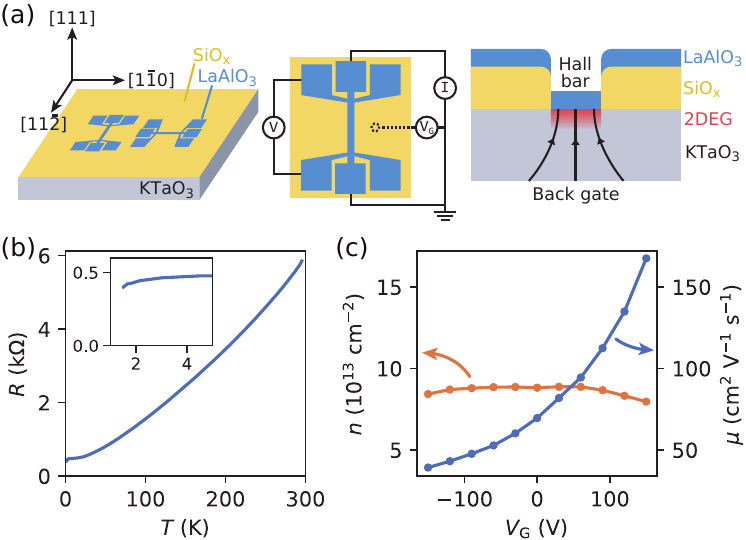}
    \caption{\textbf{(a)} Global back gate geometry. A SiO$_x$ hard-mask is used to define the Hall bars. The gate is applied to the back of the KTaO$_3$ substrate. \textbf{(b)} $R(T)$ during first cooling before applying any back-gate. \textbf{(c)} Mobility and carrier density vs back-gate at 10~K. Note that the gate range is much larger in voltage. In this geometry the gate is applied across a 0.5~mm substrate, which is why much larger voltages are required to reach similar effective fields as the locally gated sample shown in the main text.}
    \label{fig:KT029_panel1}
\end{figure}
Here we report measurements and analysis for a different sample (Sample 2) which was tuned making use of a ``global'' backgate (i.e. the traditional backgating experiment). The overall tuning of the electrostatic properties is quite similar to the the one we reported for the sample in the main text (Sample 1), as shown in Figure \ref{fig:KT029_panel1}. 

\begin{figure}[tbp]
    \centering
    \includegraphics[width = 16cm]{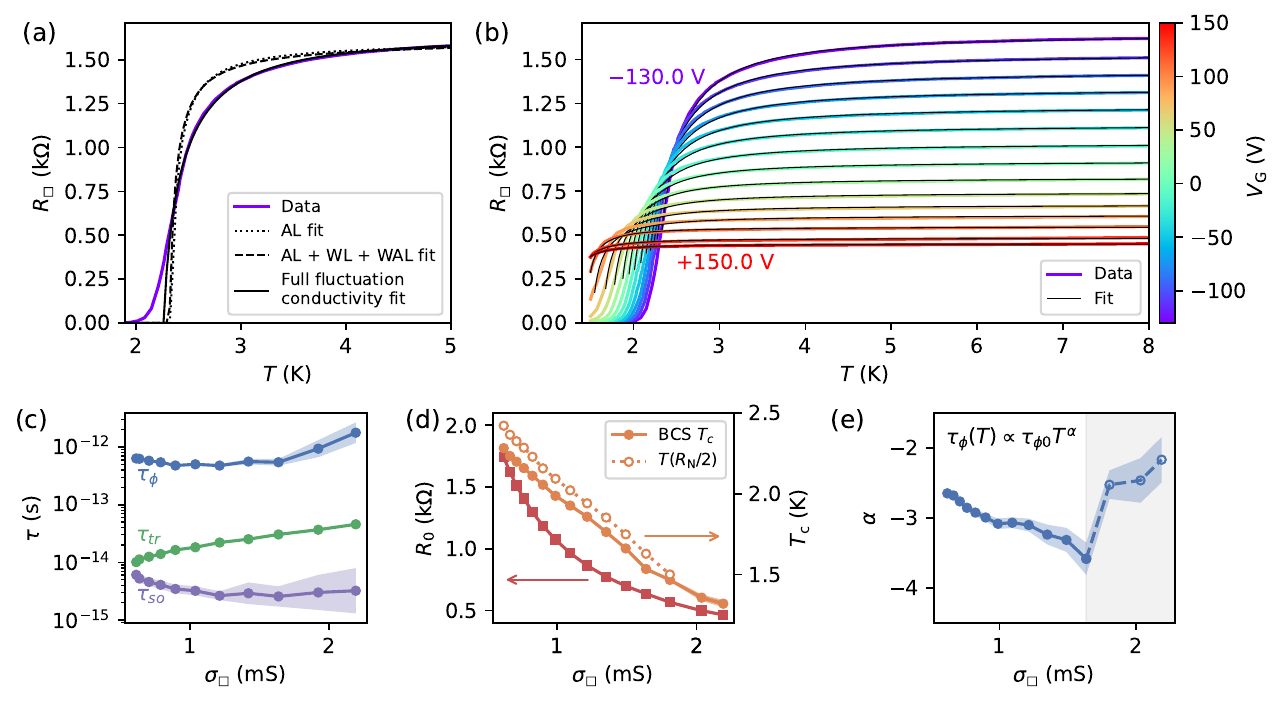}
    \caption{\textbf{Superconducting fluctuations fitting. (a)} Comparison between the simple AL model and the full fluctuation conductivity model that we adopt. \textbf{(b)} fitting of the full fluctuation model as a function of gate voltage. Note that towards the most positive side of the gating range we can no longer measure the superconducting transition, so the fit results are not reliable in this range. \textbf{(c)} Scattering times as extracted from the analysis of the magnetoconductance. The fits deteriorate above 1.7~mS as indicated by the confidence intervals becoming very broad. \textbf{(e)} Temperature exponent of the phase scattering time. The gray vertical band represents the range of fitting parameters that should be considered unreliable.}
    \label{fig:KT029_panel2}
\end{figure}

We apply the same fitting procedure to the dataset gathered for Sample 2 with the results shown in Figure~\ref{fig:KT029_panel2}. The fitting procedure here deteriorates above 75~V (as it is indicated by the gray band in panels \ref{fig:KT029_panel2}c, \ref{fig:KT029_panel2}e, \ref{fig:KT029_panel2}e) both due to an increasing uncertainty in the scattering times and to absence of a full resistive transition. 
For $V_\mathrm{G}<75$~V we find that the results are compatible with the ones that are reported in the main text:
\begin{itemize}
\item The full fluctuation conductivity model allows us to capture the temperature dependence of the resistive transition much better than the simple AL model (Figure \ref{fig:KT029_panel2}a).
\item The fitting works well for most of the gating range (\ref{fig:KT029_panel2}b)
\item The critical temperature decreases with increased conductivity.
\item We find an enhanced temperature exponent $|\alpha|>1$, which points to an electron dephasing mechanism that cannot be explained by electron-electron interference but rather electron-phonon scattering.
\end{itemize}

\begin{figure}[b]
    \centering
    \includegraphics[width=\linewidth]{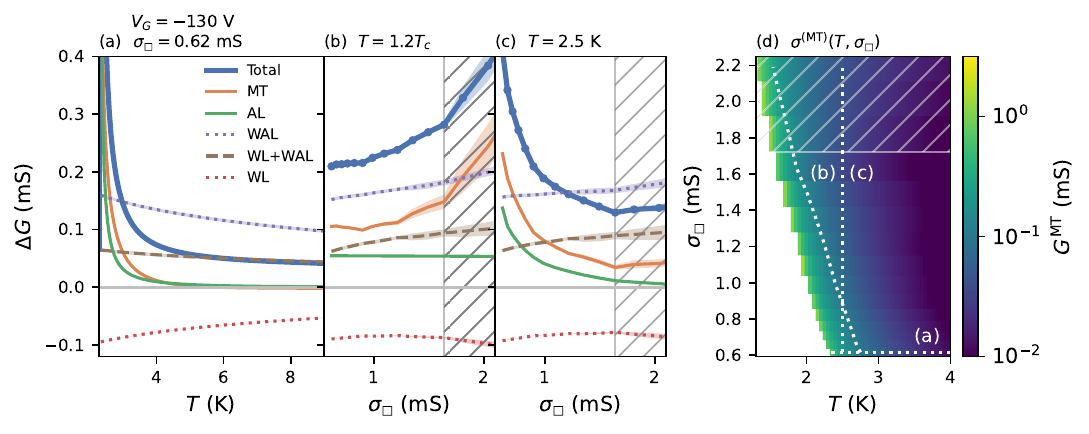}
    \caption{\textbf{Components of Fluctuation conductivity. (a)} Comparison of different contributions conductivity versus temperature at $-130$~V. Due to the long phase coherence time, weak-localization corrections are dominant at high temperatures, and Maki-Thompson corrections dominate near $T_\mathrm{c}$. \textbf{(b-c)} Comparison of the different contributions versus normal state conductance (controlled by gate voltage) at \textbf{(b)} fixed reduced temperature and \textbf{(c)} fixed temperature. At high conductance (high gate voltage) indicated by the grey region, the results are no longer reliable as the superconducting transition is no longer observable (see Fig~\ref{fig:KT029_panel2}). \textbf{(d)} Map of Maki-Thompson contributions versus temperature and normal state conductance, with lines indicating the plots in (a), (b), and (c).}
    \label{fig:KT029_panel3}
\end{figure}

Finally, when the individual components of the superconducting fluctuations are separated, we observe that the Maki Thompson contributions dominate the temperature dependence for $T<2T_\mathrm{c}$ while, for $T>2.5T_\mathrm{c}$ the temperature dependence of the resistivity is controlled by normal state contributions coming from the weak localization and anti-localization (Figure~\ref{fig:KT029_panel3}a).
When the gating dependence at constant temperature is considered we also observe that, among the superconducting contributions, MT is the most prominent. Here as well we observe a significant electrostatic tuning of the MT contribution amplitude, with a similar non-monotonic `v' shape dependence reported for Sample 1 (Figure~\ref{fig:KT029_panel3}b). Finally we observe that DCR and DOS contributions are negligible across the whole gating and temperature range.


\FloatBarrier
\section{Model considerations}

\subsection{Convergence testing}
\FloatBarrier

\begin{figure}[ht]
    \centering
    \includegraphics[width=0.5\linewidth]{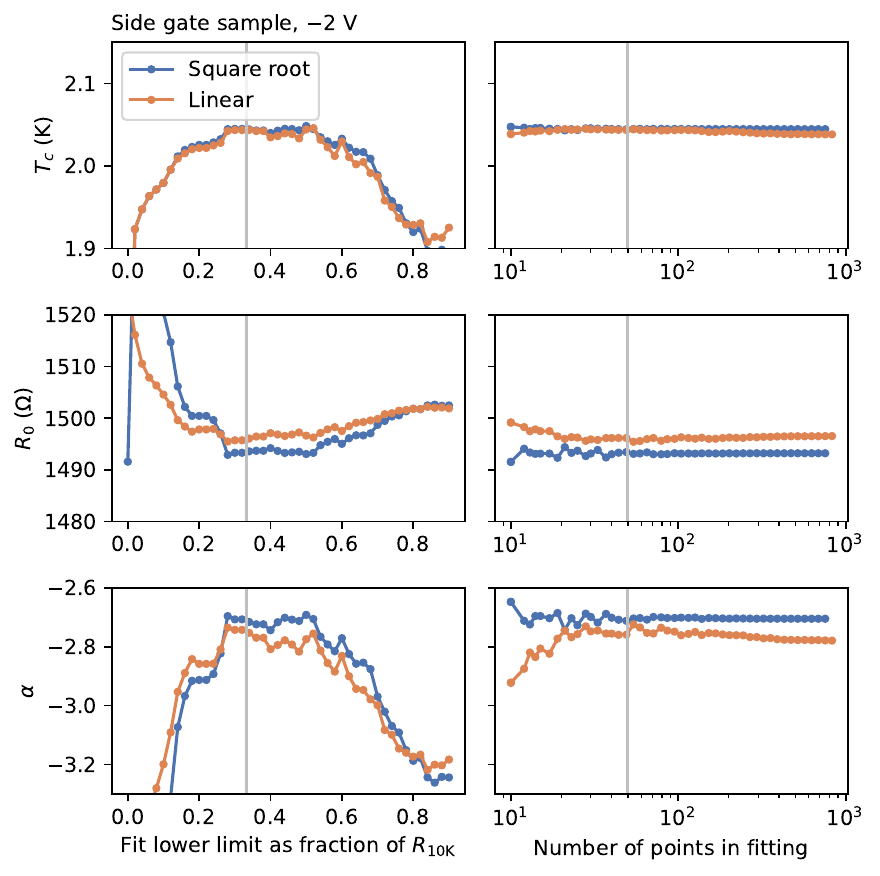}
    \caption{Convergence testing of fit results as a function of the lower limit in $T$ (left) and number of points in the fit (right), both for linearly spaced points in $T$ and square-root spaced points in $T$. Grey vertical lines indicate the final values used for the fits, along with square-root spacing.}
    \label{fig:KT027_convergence}
\end{figure}

Fit results depend on the choice of points used in the fitting. Sample inhomogeneity and BKT-like fluctuations create a `foot' - a resistive slope at near the completion of the superconducting transition which is not included in our model. To prevent this from affecting the fit accuracy, we exclude these points by only fitting above a certain fraction of $R_\square (10~\mathrm{K})$. Thus there is a trade-off between excluding this foot, but including as much of the transition as possible. To ensure we use the optimal values we fit to many cut-off points, shown in Fig~\ref{fig:KT027_convergence}. The value chosen, 1/3, is in a stable region, indicating neither the foot of the transition or lack of transition are negatively affecting the fit. We also check the number of points required for a stable result, also shown in Fig~\ref{fig:KT027_convergence}, which shows surprisingly few are required in this case. We compare both linear spacing of points in the fit, and square-root spacing of points which generates a more precise fit near the transition itself. The change in the fit results is negligilbe compared to other uncertainties, namely the electron effective mass.

For the rest of the paper, the fits use 50 points with square-root spacing and a lower cut-off of $R_\square (10~\mathrm{K})/3$.

\FloatBarrier
\subsection{Assumptions on scattering times}
\FloatBarrier

In the main text we assume that the impurity scattering time ($\tau_\mathrm{tr}$) is constant. This assumption is generally true at low temperatures. However the dielectric constant of KTaO$_3$ has a sharp increase at low temperatures. Lawless \cite{Lawless1977} reports a dielectric constant that changes between 10~K and 4~K (below which $\epsilon_r$ saturates). We believe the change $\Delta \epsilon/\epsilon\approx 5\%$ is probably not enough to change the dielectric screening significantly.

We assume a temperature independent spin orbit scattering time ($\tau_\mathrm{so}$), which is valid in many systems including LaAlO$_3$/SrTiO$_3$ \cite{stornaiuolo_weak_2014}. Recent studies show evidence of a temperature dependent $\tau_\mathrm{so}$ in Al$_2$O$_3$/KTaO$_3$ structures \cite{qin_anomalous_2025}. If we include a similar $\tau_\mathrm{so}$ temperature dependence (approximated with $T^{-1/2}$) we naturally see a stronger WAL contribution when decreasing temperature. However, since $\tau_\mathrm{so}$ doesn't directly affect the fluctuation contributions, and the localization contributions depend logarithmically on the spin-orbit time, the result of the fits is largely the same, with $T_\mathrm{c}$ and $R_0$ changing by less than 1\% and the temperature exponent of $\tau_\phi$ changing by 2\%.

There are scattering mechanisms we have not considered. There are several temperature independent mechanisms that are believed to cause electron dephasing including scattering from magnetic impurities (spin-flip scattering), random Berry phase due to the sample's roughness \cite{Mathur2001} and inhomogeneities in the doping layers \cite{Minkov2001}. 
These mechanisms are often very weak, but provide a cutoff to the otherwise diverging electron coherence times and account for the low temperature saturation of $\tau_\phi$ that is often reported \cite{Studenikin2003} \cite{Bergmann1984}. 

There have been a several works which consider electron-superconducting fluctuation scattering near $T_\mathrm{c}$ \cite{Brenig1985, Brenig1986}, which we neglect. These works find a complex temperature dependence proportional to $T$ far above the critical temperature and  $T/(\beta+\log(T/T_\mathrm{c}))$ near the critical temperature.

\FloatBarrier
\subsection{Transition broadening from inhomogeneity}
\FloatBarrier
\begin{figure}[ht]
    \centering
    \includegraphics[width=0.8\linewidth]{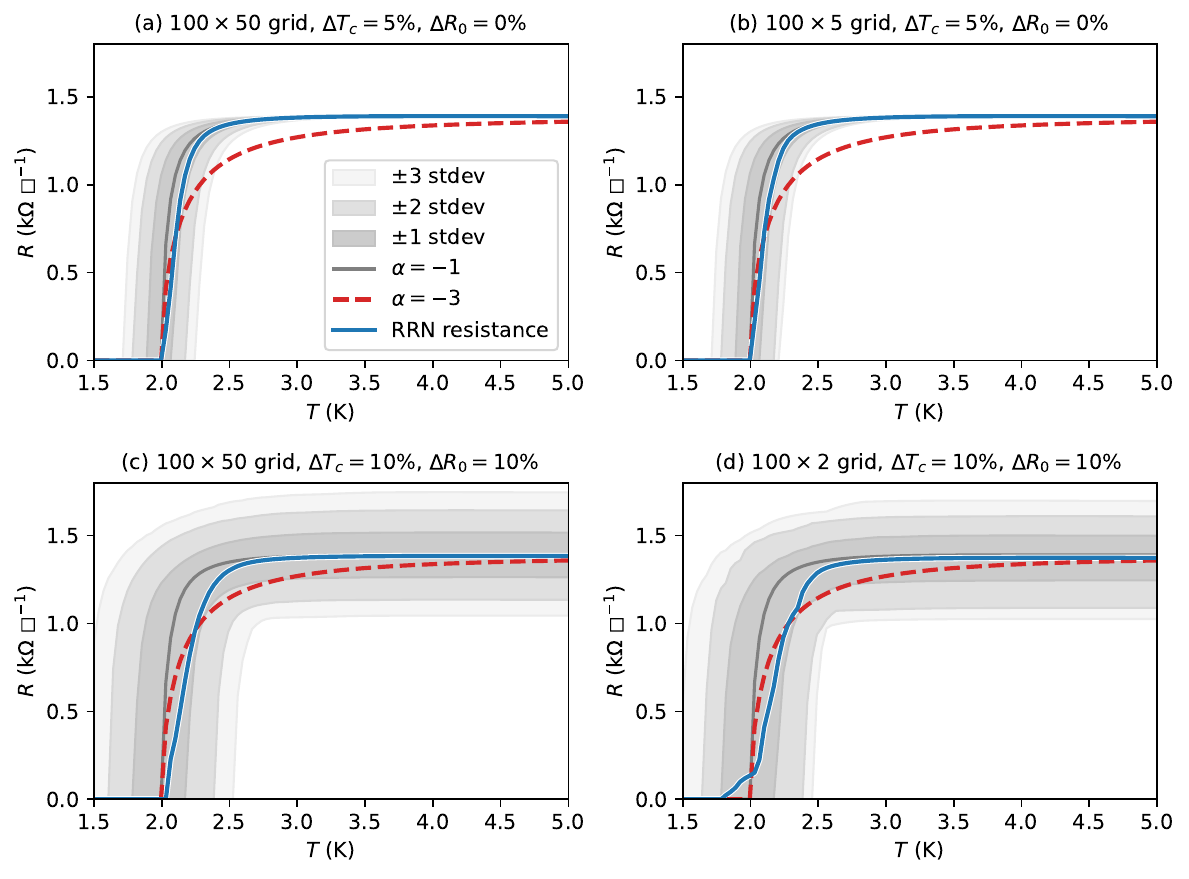}
    \caption{Random resistor network (RRN) model. \textbf{(a)}-\textbf{(d)} Plot titles describe grid size and spread in parameters. Inhomogeneity broadens the transition and decreases the `slope', but without unrealistically large $T_\mathrm{c}$ enhancements we cannot reproduce the experimentally observed broadening particularly at temperatures far above $T_\mathrm{c}$.}
    \label{fig:RRN}
\end{figure}

An important question is whether the rounding of the superconducting transition could be caused by sample inhomogeneity. In particular, we are extracting a strong temperature dependence of $\tau_\phi \approx \tau_{\phi0} T ^{-3}$ which results from the broad transition and strong excess conductivity. It is possible that our system has a `normal' electron-electron dephasing temperature dependence $\tau_\phi \approx \tau_{\phi0} T ^{-1}$ and the apparent broadening and excess conductivity results from spatially inhomogeneous critical temperatures \cite{Maccari2017}. To test this, we model an inhomogeneous system as a random resistor network.

The random resistor network is a square grid of $n_x \times n_y$ nodes each with an associated voltage, and with resistors between each node horizontally and vertically. The resistors are given an $R(T)$ with a gaussian distribution of $T_\mathrm{c}$ and/or $R_0$. Note that we do not consider phase or critical currents here, superconductors are modeled as classical resistors with infinitesimally small resistance below $T_\mathrm{c}$. For every $T$, the all resistances are calculated, a fixed voltage of 1 and 0 is applied on the left and right sides of the junction, we iteratively solve for the voltage distribution using Kirchoff's laws until convergence. Knowing the voltage of each node and resistances connecting them, we can calculate the total current and extract an effective resistance of the entire structure.

We try several configurations, two are shown in Figure~\ref{fig:RRN}. Trying different grid sizes we find it is possible to broaden the transition near $T_\mathrm{c}$. For narrow grids where low $T_\mathrm{c}$ regions are more likely to appear in series, a `foot' can appear at the bottom of the transition. However we are never able to reproduce the large, broad excess conductivity, particular at temperatures far above $T_\mathrm{c}$, without including unrealistically large $T_\mathrm{c}$ variations. The device in Figure~\ref{fig:KT029_panel2} could show more inhomogeneity resulting in broader transitions, giving us the larger extracted $\alpha$. We therefore conclude that while inhomogeneity can certainly convolute the problem and affects the accuracy of $\alpha$ extraction, we do not believe it dominates the temperature dependence.

\FloatBarrier
\subsection{Comparison to analytical first order Maki-Thompson correction}
\FloatBarrier

The first order Maki-Thompson correction close to $T_\mathrm{c}$ is given by \cite{reizer_fluctuation_1992,thompson_microwave_1970}:

\[
\sigma^{\mathrm{(MT)}} \propto \frac{2 \pi k_\mathrm{B} T}{\hbar\tau_\mathrm{GL}^{-1} - \hbar\tau_\phi^{-1}}\ln \left(\frac{\tau_\phi}{\tau_\mathrm{GL}}\right)
\]

With the Ginzburg-Landau dynamical time as

\[
\frac{\pi}{8}\hbar\tau_\mathrm{GL}^{-1} = k_\mathrm{B}T \ln\left(\frac{T}{T_\mathrm{c}}\right)
\]

\[
  \sigma^\mathrm{(MT)} \propto \frac{2 \pi^2 k_\mathrm{B} T}{8 k_\mathrm{B} T \ln{(T/T_c)}-\frac{\pi\hbar}{\tau_{\phi}}}\ln\left(\ln{(T/T_\mathrm{c})}\frac{8k_\mathrm{B}T\tau_{\phi}}{\pi\hbar}\right)
\]

At a fixed reduced temperature $t = T/T_\mathrm{c}$:


\[
  \sigma^\mathrm{(MT)}(\tau_\phi) \propto \left(8  \ln{(t)}-\frac{\pi\hbar}{k_\mathrm{B} t T_\mathrm{c} \tau_{\phi}}\right)^{-1}\ln\left(\frac{8k_\mathrm{B}tT_\mathrm{c}\tau_{\phi}}{\pi\hbar}\ln{(t)}\right)
\]

\begin{figure}[ht]
    \centering
    \includegraphics[width=0.5\linewidth]{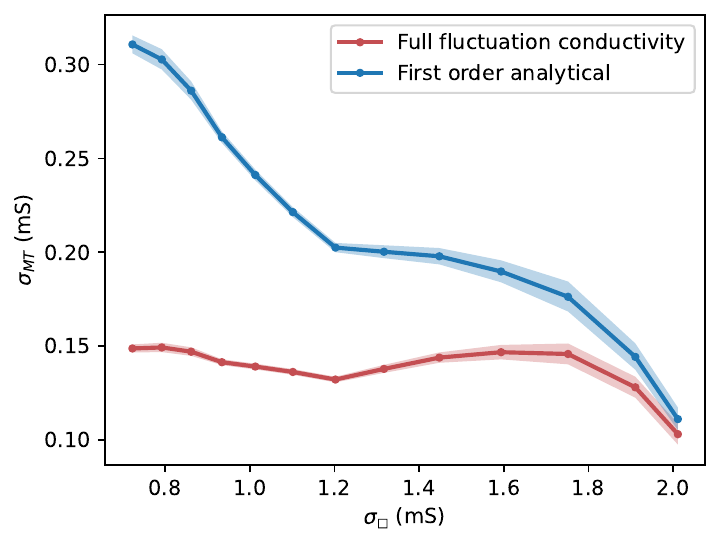}
    \caption{Comparing Maki-Thomspon from full fluctuation conducutivity and analytical first order expression at fixed reduced temperature $T/T_\mathrm{c}=1.25$, assuming $\alpha=-2.5$ and spin singlet pairing.}
    \label{fig:analytical-mt}
\end{figure}

This expression assumes only spin-singlet pairing, and hence does not depend on a spin-orbit scale or time. Comparing to the full fluctuation fit (Figure~\ref{fig:analytical-mt}) we see that the general trend is similar, but there is a significant deviation captured by the higher order corrections.

\FloatBarrier
\section{Local back-gating geometry}
\FloatBarrier

\begin{figure}[ht]
    \centering
    \includegraphics[width = 16cm]{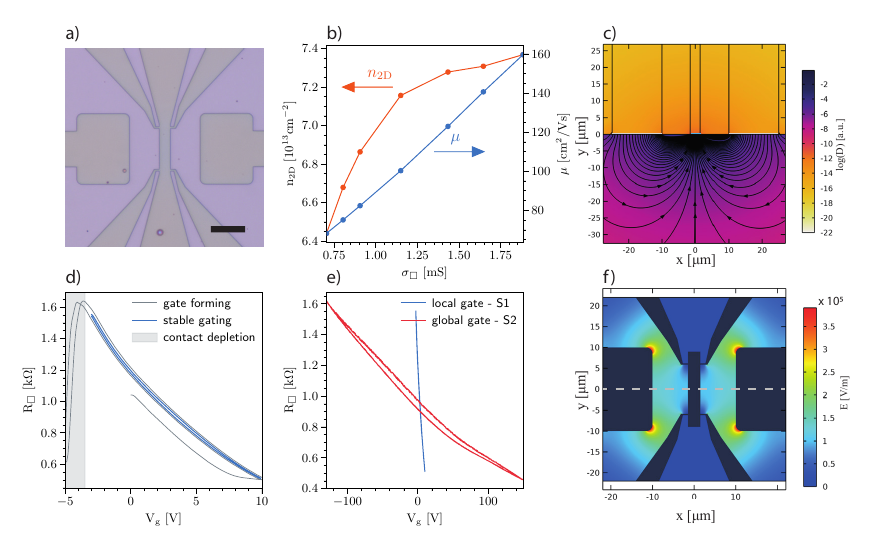}
    \caption{\textbf{Local back-gating (a)} Optical picture of the device under study right after e-beam lithography (Optical contrast is provided by PMMA on the sample). Scalebar is 10$\mu$m. \textbf{(b)} Tunability of the electronic properties of the interface. \textbf{(c)} Finite element simulation of the $\mathbf{D}$ field over a cross section of the device (see dashed line in panel f). The solid blue line highlights the position of the Hall bar while the light gray solid lines indicate the position of the gating pads. \textbf{(d)} In light gray, the gate forming curve of Sample 1. After multiple cycles between $-2$~V and 10~V the sample shows non hysteretic and stable gating. \textbf{(e)} Comparison of local and global back-gated samples showcases the enhancement of the electric-field tuning efficacy as well as a mitigation of the hysteresis of the loop. \textbf{(f)} Finite element simulation of the electric field distribution in the plane of the device (device held at 0~V, gates at 1~V).}
    \label{fig:supp_localBG}
\end{figure}

Since the first superconducting KTaO$_3$ samples were studied, it was apparent that electrostatic gating seemed to be less effective at tuning the electrical properties of the gas, especially when compared with the more studied SrTiO$_3$ based electron gasses \cite{Caviglia2008, Monteiro2017, chen_electric_2021}.
When compared with SrTiO$_3$, the gating is more challenging because of a reduced dielectric constant, $\epsilon_{\mathrm{KTO}}\approx 5\times 10^3$ \cite{Lawless1977} vs $\epsilon_{\mathrm{STO}}\approx 2\times 10^4$ \cite{Christen1994}, and a slightly higher carrier density. 
Although gating with KTaO$_3$ samples has been explored in several works\cite{chen_electric_2021, liu_tunable_2023, mallik_superfluid_2022}, highly tunable devices need thinner fragile substrate and/or the use high voltages ($\pm300$~V) which poses challenges in the cryogenic wiring and in the reproducibility of the gating experiments. 
In all ``global'' back-gating experiments (i.e. traditional gating) the samples are pasted with the aid of some type of silver paint to ensure a good thermal and electrical connection of the sample to a metallic back plate. When the gating is performed, the defective interface between silver-paint and the sample together with the high voltage causes strong hysteresis and instability in the initial stage of the gating experiments.

To overcome these challenges, inspired by previous works \cite{stornaiuolo_weak_2014,Cheng2015,Monteiro2017}, we designed a device with a ``local'' back-gate geometry (see Figure~\ref{fig:supp_localBG}a) which provides several practical advantages
\begin{enumerate}
    \item The electrodes are very close to the device under study $\approx 10~$\textmu m instead of $500~$\textmu m (the thickness of the substrates) which gives us a 50-fold enhancement of the electric field a the sample for the same gate voltage applied.
    \item Due to the lower voltages required to attain the same field, there is no need for high-voltage compliant wiring and filtering at cryogenic temperatures.
    \item The local back-gate can be patterned at the same time as the hall bar devices without the need of further fabrication steps. 
    \item The dielectric in between the local back-gate and the device is crystalline KTaO$_3$ which is more stable and more reliable than the silver paint-substrate dielectric in a global back-gate geometry. 
\end{enumerate}

We find that the local back-gating geometry does indeed provide a great enhancement of the electrostatic tuning efficacy (see Figure~\ref{fig:supp_localBG}e). However, the gating range is still limited on the positive side by the saturation of the sample's resistance and on the negative side by the depletion of the contacts (highlighted by a gray band in Figure~\ref{fig:supp_localBG}d). As it can be seen in Figure~\ref{fig:supp_localBG}f, the corners of the gating pads cause the highest electric field at the voltage probes of the device causing the aforementioned limitations.
Despite the voltage being applied on the sides of the device, most of the field lines are confined in the substrate due to the high dielectric constant of KTaO$_3$ at low temperatures (see Figure~\ref{fig:supp_localBG}).

\FloatBarrier

\bibliography{refs.bib, uldorefs.bib}